\documentclass[a4paper,fleqn,usenatbib]{mnras}
\usepackage{amsmath}
\usepackage{graphicx}	
\usepackage{amssymb}	

\newcommand{\e}{\end{equation}}
\newcommand{\bear}{\begin{eqnarray}}
\newcommand{\ear}{\end{eqnarray}}

\def\aj{AJ}
\def\apj{ApJ}
\def\apjs{ApJS}
\def\jcap{JCAP}
\def\mnras{MNRAS}
\def\aap{A\&A}
\def\prd{Physical Review D}
\def\nat{Nature}
\def\apjs{ApJS}
\def\apjl{ApJ Letters}
\def\physrep {Physics Reports}

\title[Homogeneity and periodicity in the LRG distribution] {Probing
  large scale homogeneity and periodicity in the LRG distribution
  using Shannon entropy}

\author[Pandey, B. and Sarkar, S.]  {Biswajit
  Pandey\thanks{E-mail: biswap@visva-bharati.ac.in} and Suman
  Sarkar\thanks{E-mail:suman2reach@gmail.com} \\
  Department of Physics, Visva-Bharati University, Santiniketan,
  Birbhum, 731235, India\\ }
       
 \date{\today}

 \pubyear{2015}
  
\begin{document}
\label{firstpage}
\pagerange{\pageref{firstpage}--\pageref{lastpage}}      
\maketitle
       
\begin{abstract}

We quantify the degree of inhomogeneity in the Luminous Red Galaxy
(LRG) distribution from the SDSS DR7 as a function of length scales by
measuring the Shannon entropy in independent and regular cubic voxels
of increasing grid sizes. We also analyze the data by carrying out
measurements in overlapping spheres and find that it suppresses
inhomogeneities by a factor of $5$ to $10$ on different length
scales. Despite the differences observed in the degree of
inhomogeneity both the methods show a decrease in inhomogeneity with
increasing length scales which eventually settle down to a plateau at
$\sim 150 \, h^{-1} \rm {Mpc}$. Considering the minuscule values of
inhomogeneity at the plateaus and their expected variations we
conclude that the LRG distribution becomes homogeneous at $150 \,
h^{-1} \rm {Mpc}$ and beyond. We also use the Kullback-Leibler
divergence as an alternative measure of inhomogeneity which reaffirms
our findings. We show that the method presented here can effectively
capture the inhomogeneity in a truly inhomogeneous distribution at all
length scales. We analyze a set of Monte Carlo simulations with
certain periodicity in their spatial distributions and find periodic
variations in their inhomogeneity which helps us to identify the
underlying regularities present in such distributions and quantify the
scale of their periodicity. We do not find any underlying regularities
in the LRG distribution within the length scales probed.

\end{abstract}


\section{Introduction}

Statistical homogeneity and isotropy of the Universe on sufficiently
large scales is a crucial assumption which simplifies the study of
large scale structure of the Universe and Cosmology as a whole. This
assumption which is known as the cosmological principle is one of the
fundamental pillars of modern cosmology. One can not derive this
principle in a strictly mathematical sense and can only verify this by
analyzing various cosmological observations and comparing them with
the theoretical predictions based on it. So far the most compelling
evidence of isotropy comes from the near uniform temperature of the
CMBR over the full sky \citep{penzias,smoot,fixsen}. This is further
supported by multitude of evidences such as the isotropy in angular
distributions of radio sources \citep{wilson,blake}, isotropy in the
X-ray background \citep{peeb93,wu,scharf}, isotropy of Gamma-ray
bursts \citep{meegan,briggs}, isotropy in the distribution of galaxies
\citep{marinoni,alonso}, isotropy in the distribution of supernovae
\citep{gupta,lin} and isotropy in the distribution of neutral hydrogen
\citep{hazra}. But merely having isotropy around us does not guarantee
homogeneity of the Universe. One can infer homogeneity from isotropy
only when it is assured around each and every point with the
hypothesis that the matter distribution is a smooth function of
position \citep{straumann, labini10}. Unfortunately even the isotropy
of the Universe around us is not certain as there are mounting
evidences in favour of significant anisotropy in the Universe. The
CMBR which is considered to be the best supporting evidence for
isotropy is not completely isotropic. Over the years many studies have
reported power asymmetries and unlikely alignments of low multipoles
\citep{schwarz1,land,hanlewis,moss,grupp,dai}. Recently
\citet{adeplanck1} reported a $3-\sigma$ deviations from isotropy in
the PLANCK data. There are other studies such as the hemispherical
asymmetry of cosmological parameters \citep{jackson,bengaly}, shape of
the luminosity function of galaxies \citep{appleby}, Hubble diagram of
Type Ia Supernovae \citep{schwarz2,kalus}, redshift-magnitude relation
of Type Ia Supernovae \citep{javanmardi} which find significant
evidence for anisotropy. As a result it is not straightforward to
infer homogeneity of the Universe by combining isotropy with the
Copernican principle.

 Most statistical tests to identify the transition scale to
 homogeneity rely on the number counts $n(<r)$ in spheres of radius $r$
 and its scaling with $r$ which is expected to scale as $\sim r^{3}$
 for a homogeneous distribution. The conditional density \citep{ hogg,
   labini11a} measures the average density in these spheres which is
 expected to flatten out beyond the scale of homogeneity. The
 multifractal analysis \citep{martinez90, coleman92, borgani95} uses
 the scaling of different moments of $n(<r)$ to characterize the scale
 of homogeneity. Some of the studies carried out so far with these
 methods on different galaxy surveys claim to have found a transition
 to homogeneity on sufficiently large scales $70-150 \, h^{-1} \rm
 {Mpc}$ \citep{martinez94, guzzo97, martinez98, bharad99, pan2000,
   kurokawa, hogg, yadav, prksh2,scrim,pandey15} whereas some studies
 claim the absence of any such transition out to scale of the survey
 \citep{coleman92, amen, labini07, labini09a, labini09b,
   labini11b}. Bulk flow measurements on large scales provide another
 important test for the large scale homogeneity. The galaxy
 distribution as revealed by different galaxy redshift surveys (SDSS,
 \citealt{york}; 2dFGRS, \citealt{colles}) indicates that the Universe
 is highly inhomogeneous and anisotropic on small scales. The observed
 deviations from homogeneity and isotropy arise from peculiar
 velocities resulting from density perturbations. Any form of bulk
 flows are expected to disappear on sufficiently large scales where
 the density fluctuations are very small. Analysis of a large number
 of peculiar velocity surveys indicate statistically significant large
 cosmic flows on scales of $100 \, h^{-1}\, {\rm Mpc}$
 \citep{watkins}. Combined studies with the WMAP data and x-ray
 cluster catalog indicates strong and coherent bulk flow out to
 $\gtrsim 300 \, h^{-1}\, {\rm Mpc}$
 \citep{kashlinsky1,kashlinsky2}. Contrary to these claims a recent
 study of low-redshift Type-Ia Supernovae suggest no excess bulk flow
 on top of what is expected in $\Lambda$CDM model \citep{huterer}. If
 large scale bulk motions really exist then they indicate that there
 are significant density fluctuations on very large scales. The
 results from these studies clearly indicate that there is no clear
 consensus in this issue yet. The presence of inhomogeneities on very
 large scales has several important consequences. Inhomogeneities,
 through the backreaction mechanism can provide an alternate
 explanation of a global cosmic acceleration without requiring any
 additional dark energy component \citep{buchert97, buchert01,
   schwarz, kolb06, paranjape09, kolb10, ellis}. Spherically symmetric
 radially inhomogeneous Lemaitre-Tolman-Bondi (LTB) models can account
 for the distant supernova data in a dust universe without requiring
 any dark energy \citep{garcia,enqvist}. A large void can also mimic
 an apparent acceleration of expansion
 \citep{tomita01,hunt}. Fortunately such models can be constrained
 with observations such as SNe, CMB, BAO and measurements of angular
 diameter distance and Hubble parameter \citep{zibin, clifton, biswas,
   chris, larena, valken}. There would be a major paradigm shift in
 cosmology if the assumption of cosmic homogeneity is ruled out with
 high statistical significance by multiple data sets and it is
 important to test the assumption of homogeneity on multiple datasets
 with different statistical tools.

\citet{pandey} introduce a method based on Shannon entropy
\citep{shannon48} for characterizing inhomogeneities and applied the
method on some Monte Carlo simulations of inhomogeneous distributions
and N-body simulations which show that the proposed method has great
potential for testing the large scale homogeneity in galaxy redshift
surveys. \citet{pandey15} applied this method to galaxy distributions
from SDSS DR12 \citep{alam} and find that the inhomogeneities in the
galaxy distributions persist at least upto a length scale of $120 \,
h^{-1}\, {\rm Mpc}$. \citet{pandey} show that the methods for testing
homogeneity based on the number counts are contaminated by confinement
and overlapping biases which result from the systematic migration and
overlap between the spheres with increasing radii. These biases which
are inevitable in any finite volume sample, artificially suppress
inhomogeneities at each scale and particularly on large
scales. Recently \citet{kraljic} use the count in cells method in Dark
Sky simulations \citep{skillman} to show the spurious effects in the
measurement of correlation dimension due to the confinement and
overlapping biases. One can circumvent these biases by using number
counts in independent spheres but the availability of only a very
small number of independent spheres at increasing radii makes the
statistics too noisy for most galaxy samples available including the
SDSS Main galaxy sample. Using count in cells method as well as new
methods based on anomalous diffusion and random walks \citet{kraljic}
show that the Millennium Run simulation \citep{springel} volume is too
small and prone to bias to reliably identify the onset of
homogeneity. We require galaxy distributions covering very large
volume to serve the purpose. The luminous red galaxy distribution
(LRG) \citep{eisen} extends to a much deeper region of the Universe as
compared to the SDSS Main galaxy sample as they can be observed to
greater distances as compared to normal $L_{\star}$ galaxies for a
given magnitude limit. Their stable colors also make them relatively
easy to pick out from the rest of the galaxies using the SDSS
multi-band photometry. The enormous volume covered by the SDSS LRG
distribution provides us an unique opportunity to test the assumption
of homogeneity on large scales with unprecedented confidence. We
modify our method \citep{pandey} so as to carry out the analysis with
non-overlapping independent regions upto sufficiently large length
scales. We also analyze the LRG data with overlapping spheres and
compare the findings with that obtained using independent regions to
asses the suppression of inhomogeneities on different length scales
due to overlapping and confinement biases.

Beside analyzing the issue of cosmic homogeneity using the LRG
distribution we also address another important issue in the present
work. Some studies reported to have found periodicity in the observed
distribution of galaxies and galaxy clusters \citep{broad,einasto}.
Both \citet{broad} and \citet{einasto} find an apparent regularity in
the galaxy distribution at a characteristic length scale of $120-130
\, h^{-1}\, {\rm Mpc}$. It has been suggested that in an oscillating
Universe the oscillation of the Hubble parameter can cause such
periodicity in the density fluctuations \citep{mori,
  hirano}. \citet{einas1} use the oscillations in the cluster
correlation function to quantify such pictures of regularity in the
galaxy distribution. The correlation function being an isotropic
statistics smooths out the anisotropic signal and may not be the best
statistics to describe such anisotropic structures. Recently
\citet{pandey2} suggest a method which shows that Shannon entropy can
be also employed to test the isotropy of a distribution. We shall use
some inhomogeneous and anisotropic distributions with regularity in
their distributions and test if the method presented here can describe
such underlying regularities in the distributions. A recent study by
\citet{ryabin} find quasi-periodical features in the distribution of
LRGs. Recently \citet{maret1} detected two peaks in the distance
distributions of rich galaxy groups and clusters at $120 \, h^{-1}\,
{\rm Mpc}$ and $240 \, h^{-1}\, {\rm Mpc}$ around the Abell cluster
A2142. Another study by \citet{maret2} find evidence for shell like
structures in the SDSS galaxy distribution where galaxy clusters have
maxima in the distance distribution to other galaxy groups and
clusters at the distance of about $120 \, h^{-1}\, {\rm
  Mpc}$. Galaxies are distributed in a complex filamentary network
which is often referred as the Cosmic web. Filaments are known to be
the largest known statistically significant coherent structures in the
Universe \citep{bharad04,pandey05,pandey10}. \citet{pandey11} find
that the filaments are statistically significant upto a length scales
of $120 \, h^{-1}\, {\rm Mpc}$ in the LRG distribution and longer
filaments, though possibly present in the data, are not statistically
significant and are the outcome of chance alignments. These studies
suggest that there may be a characteristic scale present in the galaxy
distribution other than the Baryon Acoustic Oscillations (BAO)
feature observed at $100 \, h^{-1}\, {\rm Mpc}$ \citep{eisen1}. In an
earlier study \citet{pandey15} find that the inhomogeneities in the
SDSS Main galaxy distribution persist at least upto a length scales of
$120 \, h^{-1}\, {\rm Mpc}$ and we would like to test if the
inhomogeneities in the LRG distribution exhibit any regularity on even
larger scales.

Throughout our work, we have used the flat $\Lambda$CDM cosmology with
$\Omega_m = 0.3, \; \Omega_{\Lambda} = 0.7 \mbox{ and } h = 1$.

\section {METHOD OF ANALYSIS}

\citet{pandey} propose a method based on the Shannon entropy to study
inhomogeneities in a 3D distribution. Shannon entropy
\citep{shannon48} is originally proposed by Claude Shannon to quantify
the information content in strings of text. It gives a measure of the
amount of information required to describe a random variable. The
Shannon entropy for a discrete random variable $X$ with $n$ outcomes
$\{x_{i}:i=1,....n\}$ is a measure of uncertainty denoted by $H(X)$
defined as,

\begin{equation}
H(X) =  - \sum^{n}_{i=1} \, p(x_i) \, \log \, p(x_i)
\label{eq:shannon1}
\end{equation}

where $p(x)$ is the probability distribution of the random variable
$X$.  

We modify the method proposed by \citet{pandey} which employs spheres
of varying radii $r$ to measure the number counts inside them. The
spheres are allowed to overlap which suppress inhomogeneities and the
degree of overlap gradually increases with increasing radii. In order
to prevent overlapping of the measuring volumes which hinders
measurement of inhomogeneity on large scales we embed the galaxy
distribution in a $d \, h^{-1}\, {\rm Mpc} \times d \, h^{-1}\, {\rm
  Mpc} \times d \, h^{-1}\, {\rm Mpc}$ three dimensional rectangular
grid. This divides the entire survey region into a number of regular
cubic voxels. The voxels which are beyond the boundary of the survey
region are marked and discarded from the analysis. We also discard the
voxels near the boundary which are partly inside the survey region to
avoid any edge effects. We identify the voxels which are fully inside
the survey region and only consider them in our analysis. The grid
size $d$ is a variable which can be varied within a suitable
range. Each choice of $d$ would result into a different number of
analyzable voxels and galaxies within the survey region. Let $N_{d}$
be the number of resulting voxels which lie entirely within the survey
region for a grid size $d$. We count the number of galaxies inside
each of the $N_{d}$ voxels available for each choice of grid size
$d$. Let $n_{i}$ be the number counts inside the $i^{th}$ voxel where
the voxel index $i$ runs from $1$ to $N_{d}$. In general for a
specific choice of the grid size if we randomly pick up a galaxy it
can reside only in one of the $N_{d}$ voxels available for that
$d$. But which particular voxel does it belong to ?  The answer to
this question has $N_{d}$ likely outcomes. We define a separate random
variable $X_{d}$ for each grid size $d$ which has $N_{d}$ possible
outcomes each given by, $f_{i,d}=\frac{n_{i}}{\sum^{N_{d}}_{i=1} \,
  n_{i}}$ with the constraint $\sum^{N_{d}}_{i=1} \, f_{i,d}=1$.  The
Shannon entropy associated with the random variable $X_{d}$ can be
written as,
\begin{eqnarray}
H_{d}& = &- \sum^{N_{d}}_{i=1} \, f_{i,d}\, \log\, f_{i,d} \nonumber\\ &=& 
\log(\sum^{N_{d}}_{i=1}n_{i}) - \frac {\sum^{N_{d}}_{i=1} \,
  n_i \, \log n_i}{\sum^{N_{d}}_{i=1} \, n_i}
\label{eq:shannon2}
\end{eqnarray}
Where the base of the logarithm is arbitrary and we choose it to be
$10$. $f_{i,d}$ will have the same value $\frac{1}{N_{d}}$ for all the
voxels when $n_{i}$ is same for all of them. This is an ideal
situation when each of the $N_{d}$ voxels available contain exactly
the same number of galaxies within them. This maximizes the Shannon
entropy to $(H_{d})_{max}=\log \, N_{d}$ for grid size $d$. We define
the relative Shannon entropy as the ratio of the entropy of a random
variable $X_{d}$ to the maximum possible entropy $(H_{d})_{max}$
associated with it. The relative Shannon entropy
$\frac{H_{d}}{(H_{d})_{max}}$ for any grid size $d$ quantifies the
degree of uncertainty in the knowledge of the random variable
$X_{d}$. The distribution of $f_{i,d}$ become completely uniform when
$\frac{H_{d}}{(H_{d})_{max}}=1$ is reached. Equivalently
$1-\frac{H_{d}}{(H_{d})_{max}}$ quantify the residual information and
can be treated as a measure of inhomogeneity. The fact that galaxies
are not residing in any particular voxel and rather are distributed
across the available voxels with different probabilities acts as the
source of information. If all the galaxies would have been residing in
one particular voxel leaving the rest of them as empty then there
would be no uncertainty and no information at all making $H_{d}=0$ or
$1-\frac{H_{d}}{(H_{d})_{max}}=1$. This fully determined hypothetical
situation corresponds to maximum inhomogeneity. On the other hand when
all the $N_{d}$ voxels are populated with equal probabilities it would
be most uncertain to decide which particular voxel a randomly picked
galaxy belongs to. This maximizes the information entropy to
$H_{d}=\log \, N_{d}$ turning $1-\frac{H_{d}}{(H_{d})_{max}}=0$. This
corresponds to a situation when the distribution is completely
homogeneous. The galaxy distribution is expected to be inhomogeneous on
small scales but with increasing grid size $d$ one would expect it to
be homogeneous on some scale provided the Universe is homogeneous on
large scales. It may be noted here that one may prefer to use
non-overlapping spheres to carry out the measurements but use of
regular cubic voxels allow the optimal use of the available space.

Alternatively one can measure the inhomogeneity by using the
Kullback-Leiblar(KL) divergence or information divergence. In
information theory KL divergence is used to measure the difference
between two probability distributions $p(x)$ and $q(x)$.\\
\begin{equation}
 D_{KL}(p|q)=\sum_{i} p(x_{i}) \, \log \frac{p(x_{i})}{q(x_{i})}
\label{eq:kld}
\end{equation}

Let $f_{Di}$ be the distribution corresponding to the data for which
homogeneity is to be tested and $f_{Ri}$ is the distribution for a
homogeneous Poisson random distribution where both the distributions
occupy same 3D volume with identical geometry and are represented by
same number of points. We simulate a Monte Carlo realization of a
homogeneous Poisson distribution within the survey region using same
number of points as the galaxies in that region. The Monte Carlo
sample of the homogeneous random Poisson distribution serve as our
reference distribution for measuring the information divergence of the
LRG distribution with respect to it.

The KL divergence between the actual and random data is then given
by,\\
\begin{equation}
 D_{KL}(D|R)=\frac{(\sum_{i} n_{Di}\,\log n_{Di}-\sum_{i} n_{Di}\,\log
   n_{Ri})}{\sum_{i} n_{Di}}-\log\frac{\sum_{i} n_{Di}}{\sum_{i} n_{Ri}}
\label{eq:kldiv}
\end{equation}

where $n_{Di}$ and $n_{Ri}$ are the counts in the $i^{th}$ voxel for
actual data and random data respectively.

\begin{figure*}
  \resizebox{9cm}{!}{\rotatebox{0}{\includegraphics{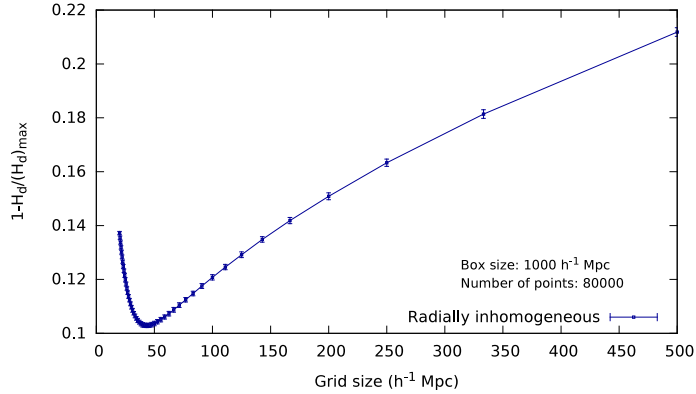}}}\\
\caption{This shows the variation of inhomogeneity
  $1-\frac{H_{d}}{(H_{d})_{max}}$ as a function of length scales in a
  radially inhomogeneous Poisson distribution where density falls of
  as $\frac{1}{r^{2}}$ from one of the corners of a cube with sides
  $1000 \, h^{-1} \, {\rm Mpc}$. The measurements are done using
  independent and regular cubic voxels as described in section 2. The
  errorbars shown here are the $1-\sigma$ variations from the $10$
  Monte Carlo realizations used.}
  \label{fig:inhfig}
\end{figure*}

\begin{figure*}
 \resizebox{7cm}{!}{\rotatebox{0}{\includegraphics{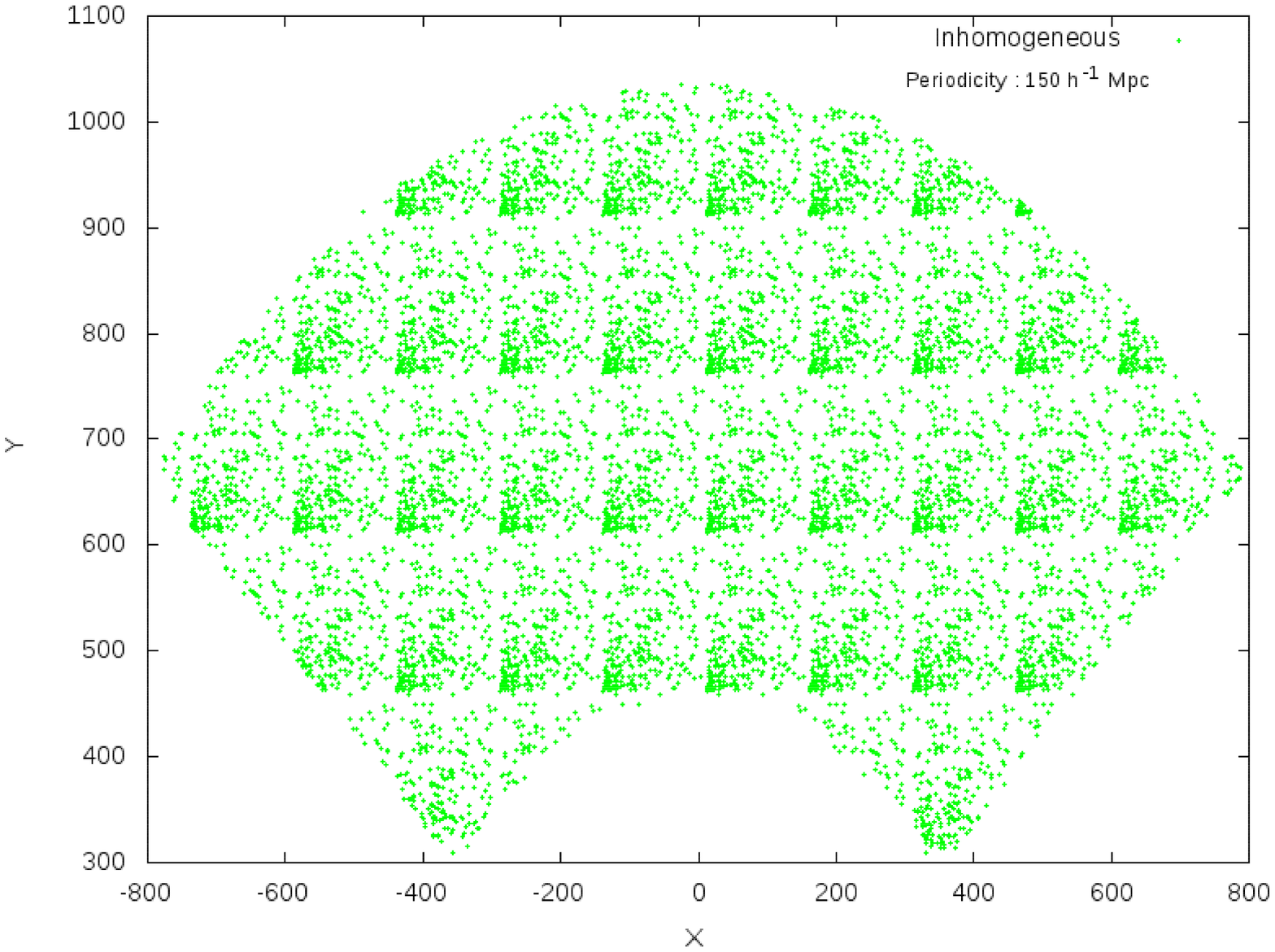}}}%
 \resizebox{7cm}{!}{\rotatebox{0}{\includegraphics{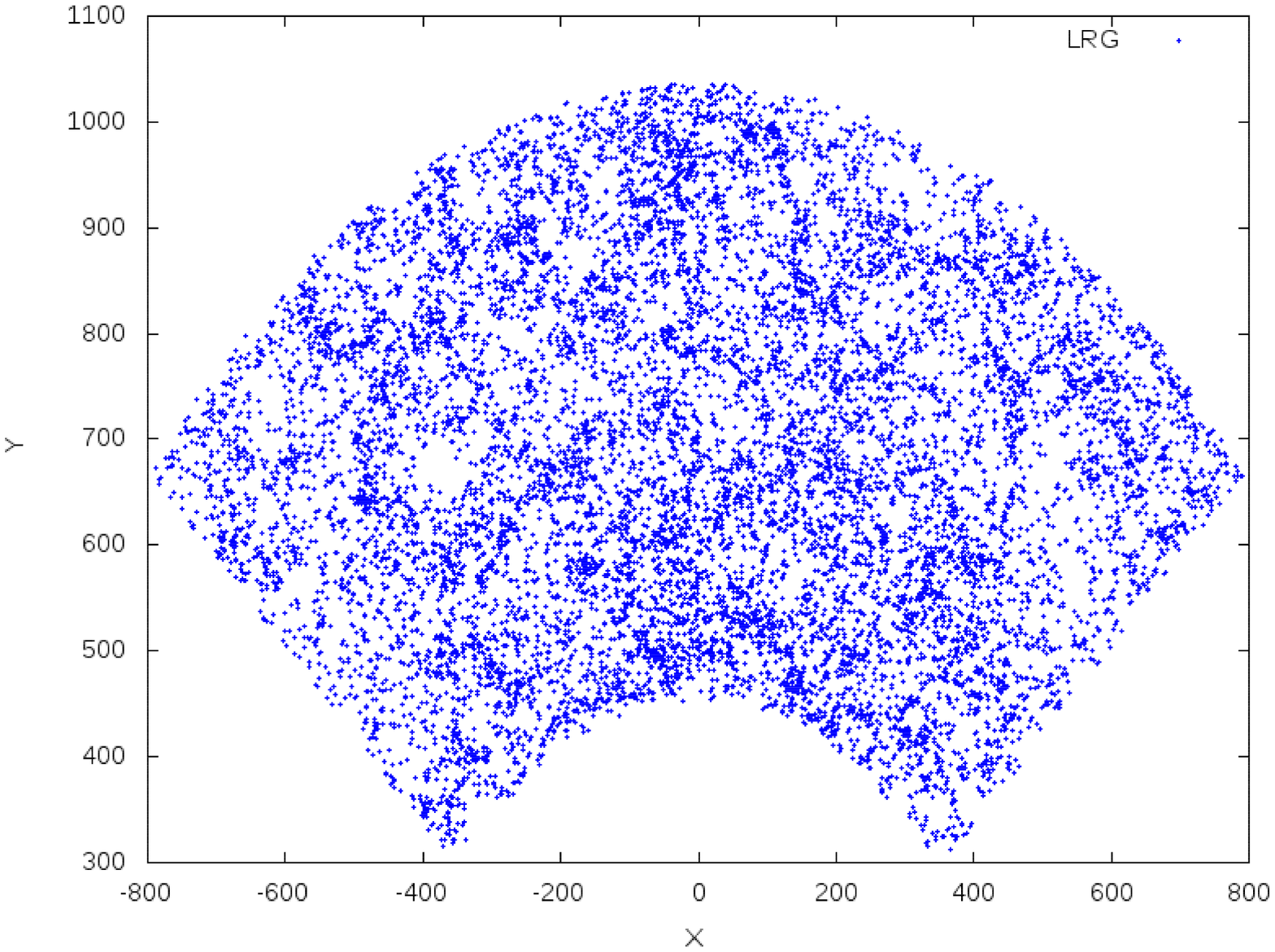}}}\\
 \caption{ The left panel shows the 2D projected view of an
   anisotropic and inhomogeneous Poisson distribution which
   periodically repeats itself at regular intervals of $150 \, h^{-1}
   \, {\rm Mpc}$. The right panel show the projected LRG
   distribution. Both distributions shown here have an uniform
   thickness of $200 \, h^{-1} \, {\rm Mpc}$.  The distributions which
   are analyzed actually have radially varying thickness.}
 \label{fig:dist}
\end{figure*}

\begin{figure*}
 \resizebox{9cm}{!}{\rotatebox{0}{\includegraphics{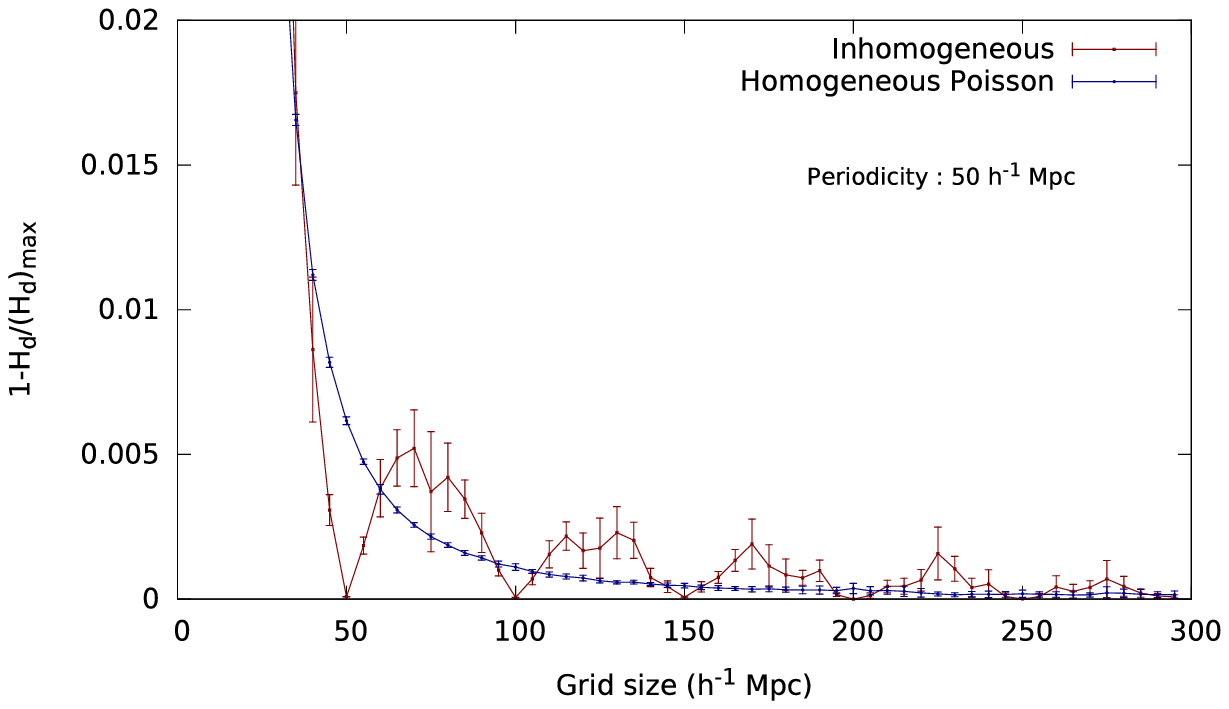}}}%
 \resizebox{9cm}{!}{\rotatebox{0}{\includegraphics{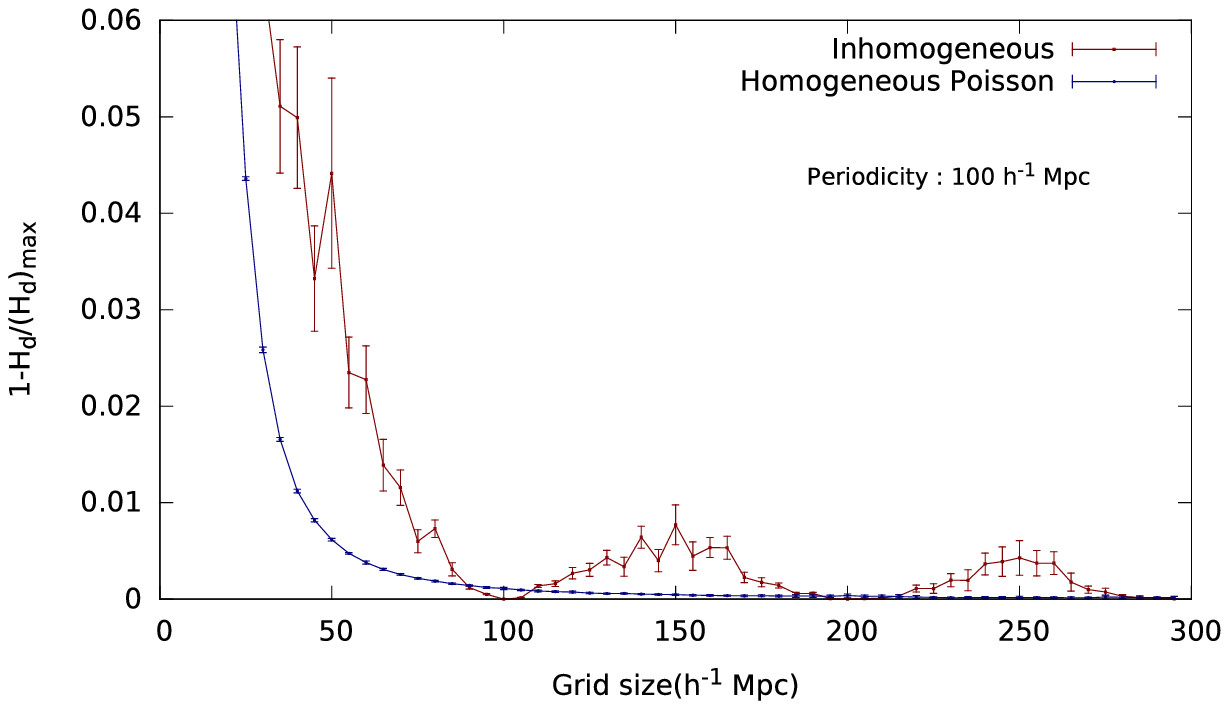}}}\\
 \resizebox{9cm}{!}{\rotatebox{0}{\includegraphics{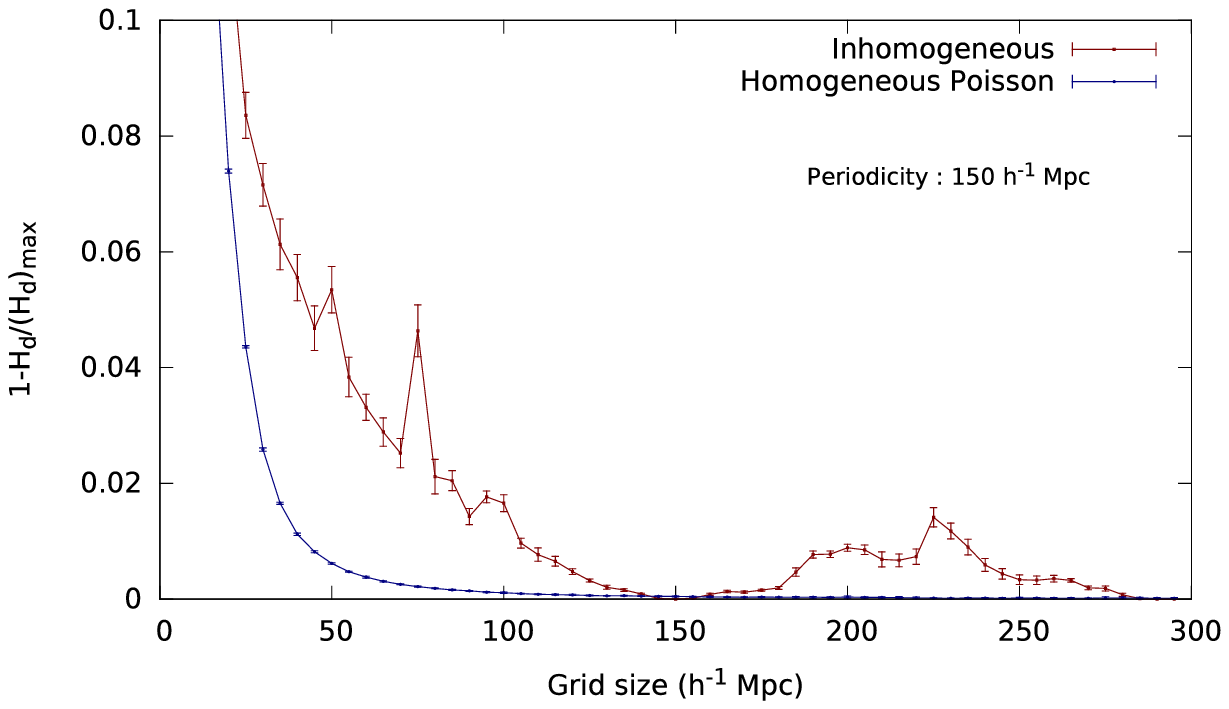}}}\\
 \caption{The top left figure shows the inhomogeneity
   $1-\frac{H_{d}}{(H_{d})_{max}}$ as a function of length scales in
   an anisotropic and inhomogeneous Poisson distribution which
   periodically repeats itself at regular intervals of $50 \, h^{-1}
   \, {\rm Mpc}$. The top right and the bottom panels show the same
   for anisotropic and inhomogeneous Poisson distributions which are
   periodic on scales of $100 \, h^{-1} \, {\rm Mpc}$ and $150 \,
   h^{-1} \, {\rm Mpc}$ respectively. The results shown in all the
   panels in this figure are obtained using independent and regular
   cubic voxels as described in section 2. The error-bars shown here
   in all the panels are the $1-\sigma$ variations from the $10$ Monte
   Carlo realizations used in each case.}
  \label{fig:pinhfig}
\end{figure*}

\begin{figure*}
 \resizebox{9cm}{!}{\rotatebox{0}{\includegraphics{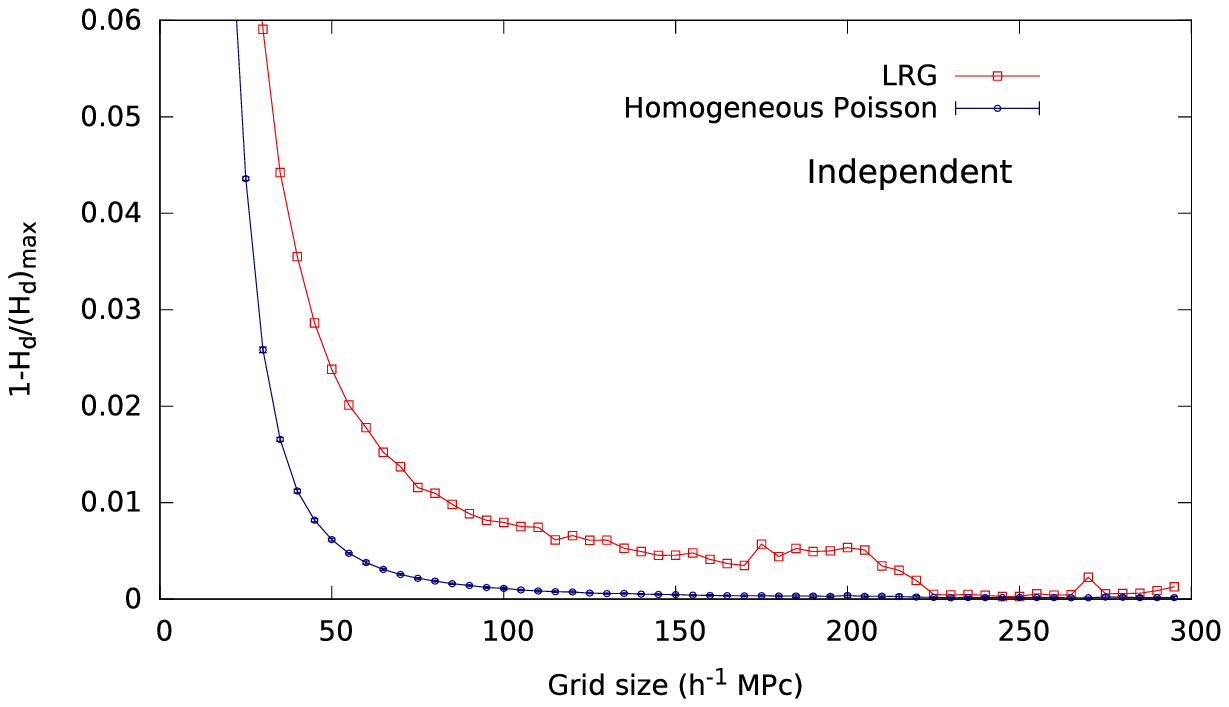}}}%
 \resizebox{9cm}{!}{\rotatebox{0}{\includegraphics{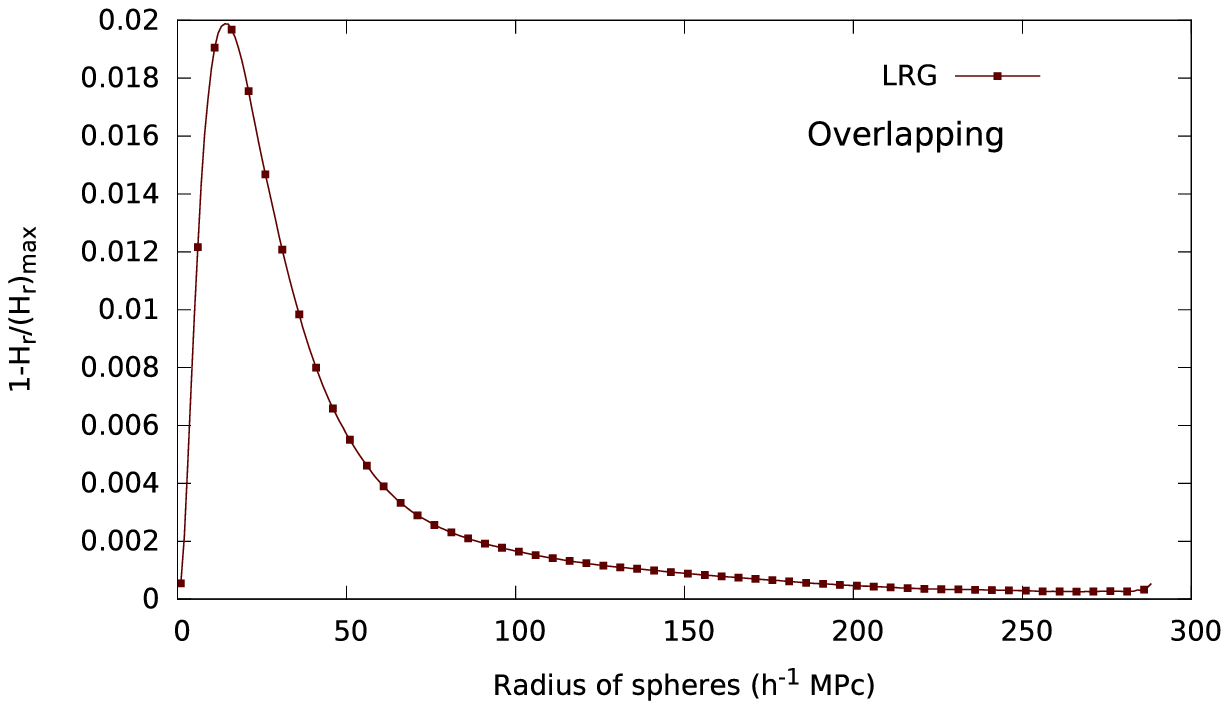}}}\\
 \resizebox{9cm}{!}{\rotatebox{0}{\includegraphics{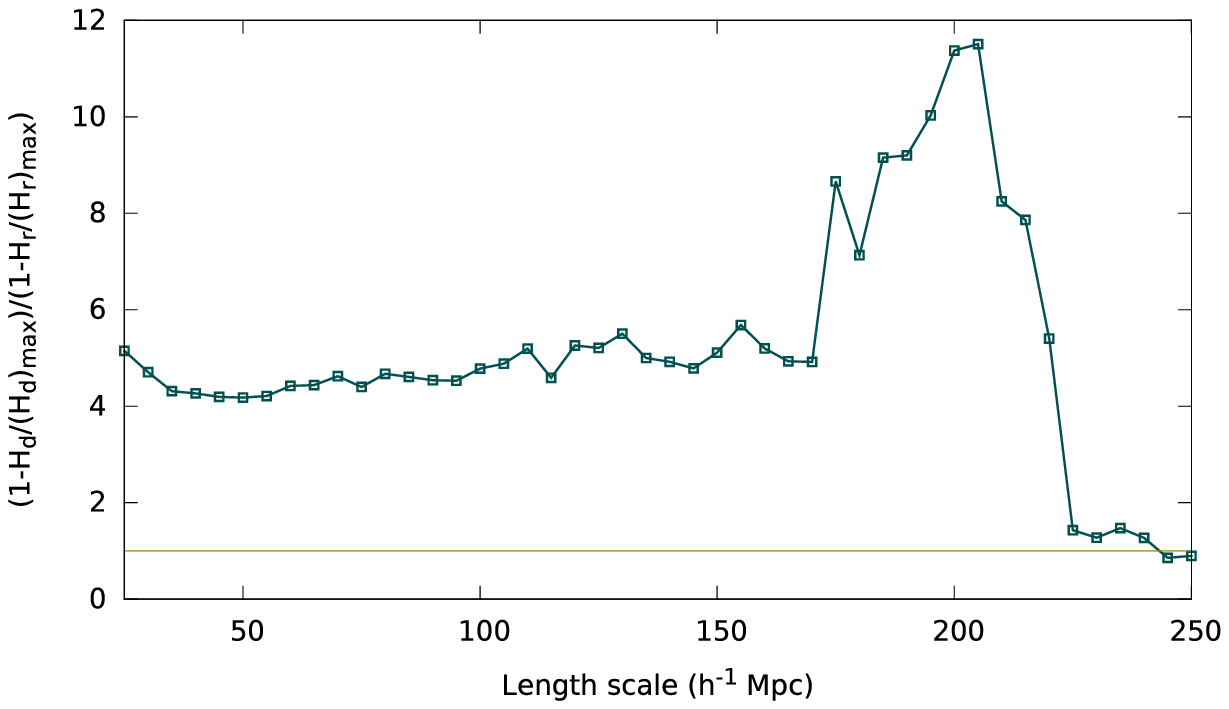}}}%
 \resizebox{9cm}{!}{\rotatebox{0}{\includegraphics{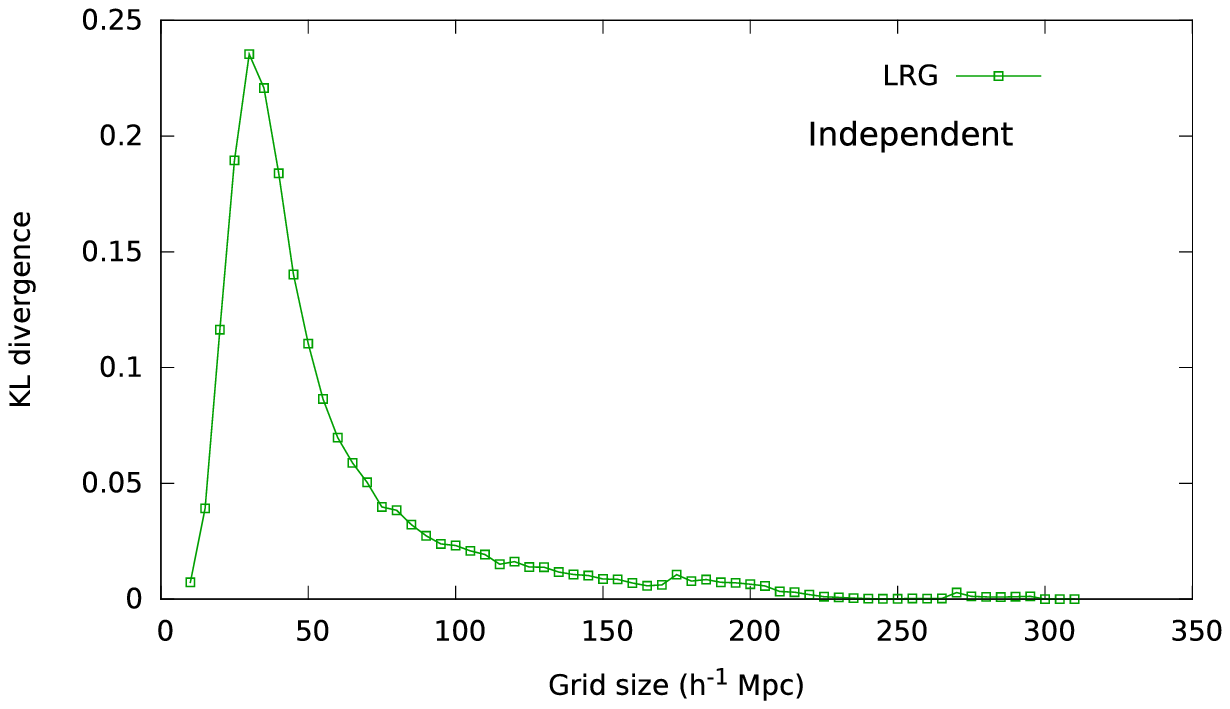}}}\\
 \caption{The top left panel shows the inhomogeneity
   $1-\frac{H_{d}}{(H_{d})_{max}}$ as function of length scales in the
   LRG distribution and a homogeneous Poisson distribution when the
   analysis is done with independent voxels as described in section
   2. The top right panel shows the inhomogeneity
   $1-\frac{H_{r}}{(H_{r})_{max}}$ as a function of $r$ in the LRG
   distribution when the analysis is carried out with overlapping
   spheres \citep{pandey,pandey15}. The bottom left panel show the
   ratio of inhomogeneity on different length scales when measured
   using independent voxels and overlapping spheres. The bottom right
   panel show the Kullback-Leibler (KL) divergence measure as a
   function of length scales when the LRG data is analyzed using
   independent voxels.}
  \label{fig:lrghfig}
\end{figure*}

\begin{figure*}
\resizebox{8cm}{!}{\rotatebox{0}{\includegraphics{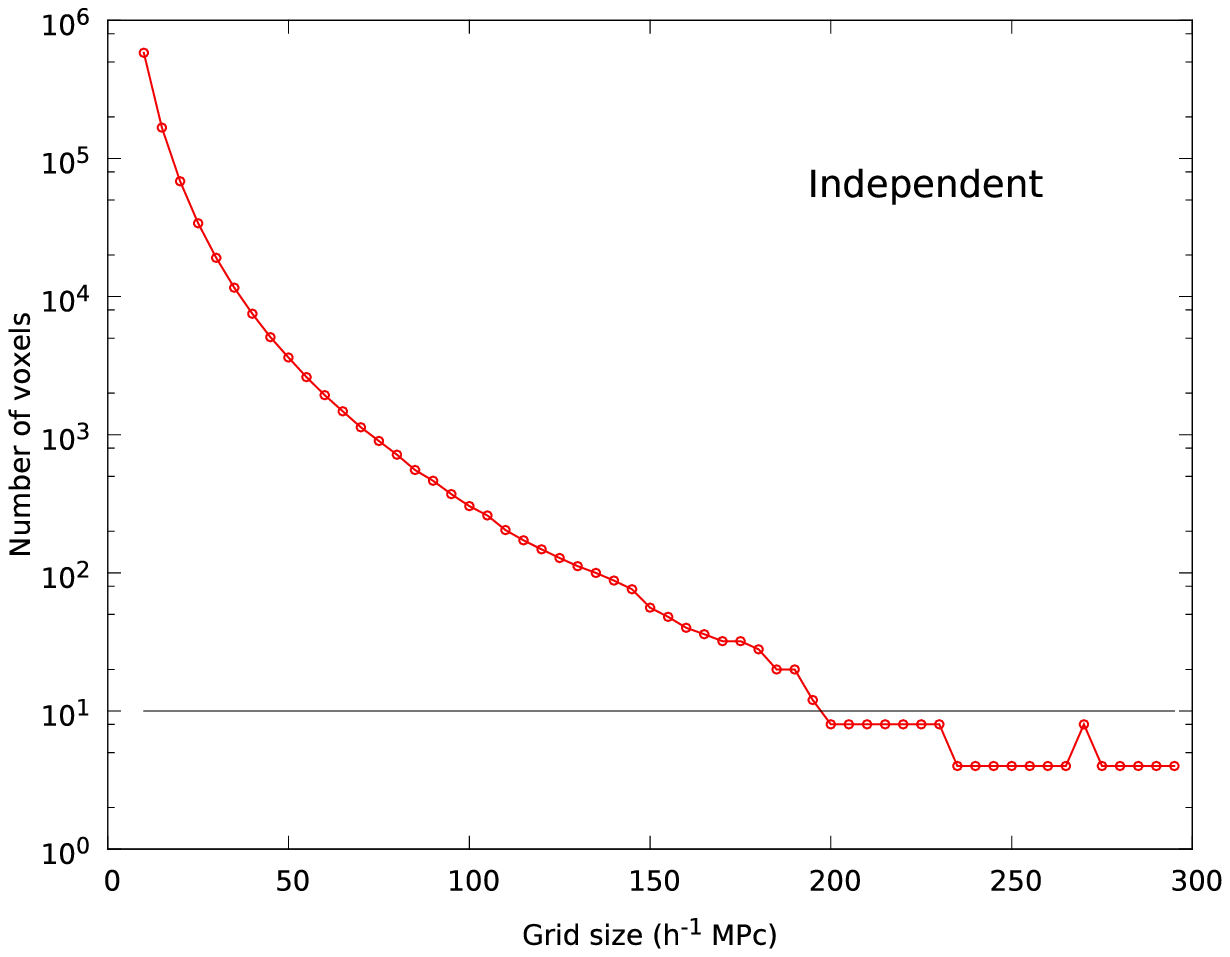}}}%
\resizebox{8cm}{!}{\rotatebox{0}{\includegraphics{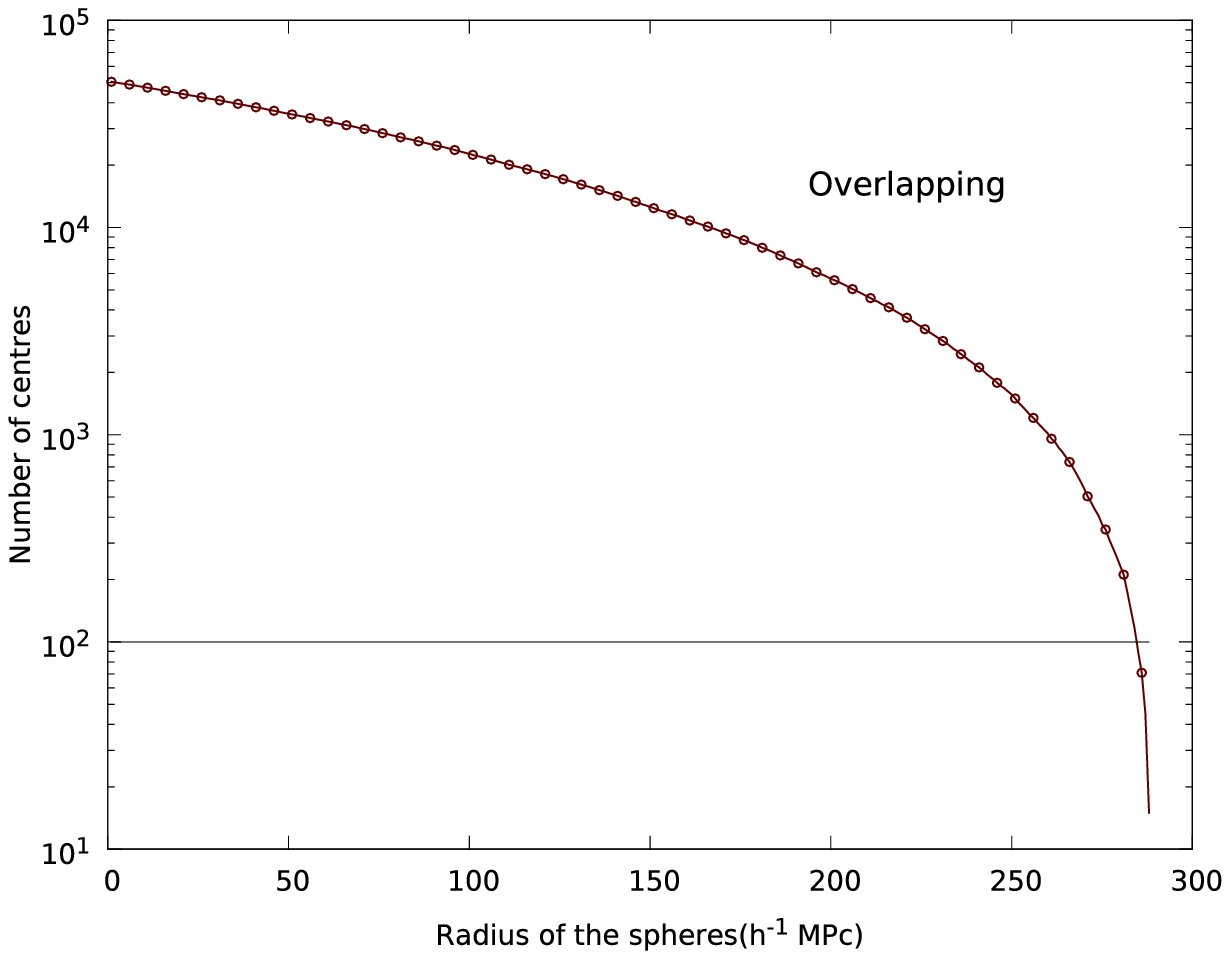}}}\\
 \caption{The left panel shows how the number of independent voxels
   available within the survey region changes with increasing grid
   sizes. The right panel shows the number of centres available at
   each radii in the LRG distribution when the analysis is carried out
   with overlapping spheres.}
  \label{fig:lrgmrfig}
\end{figure*}

\begin{figure*}
\resizebox{9cm}{!}{\rotatebox{0}{\includegraphics{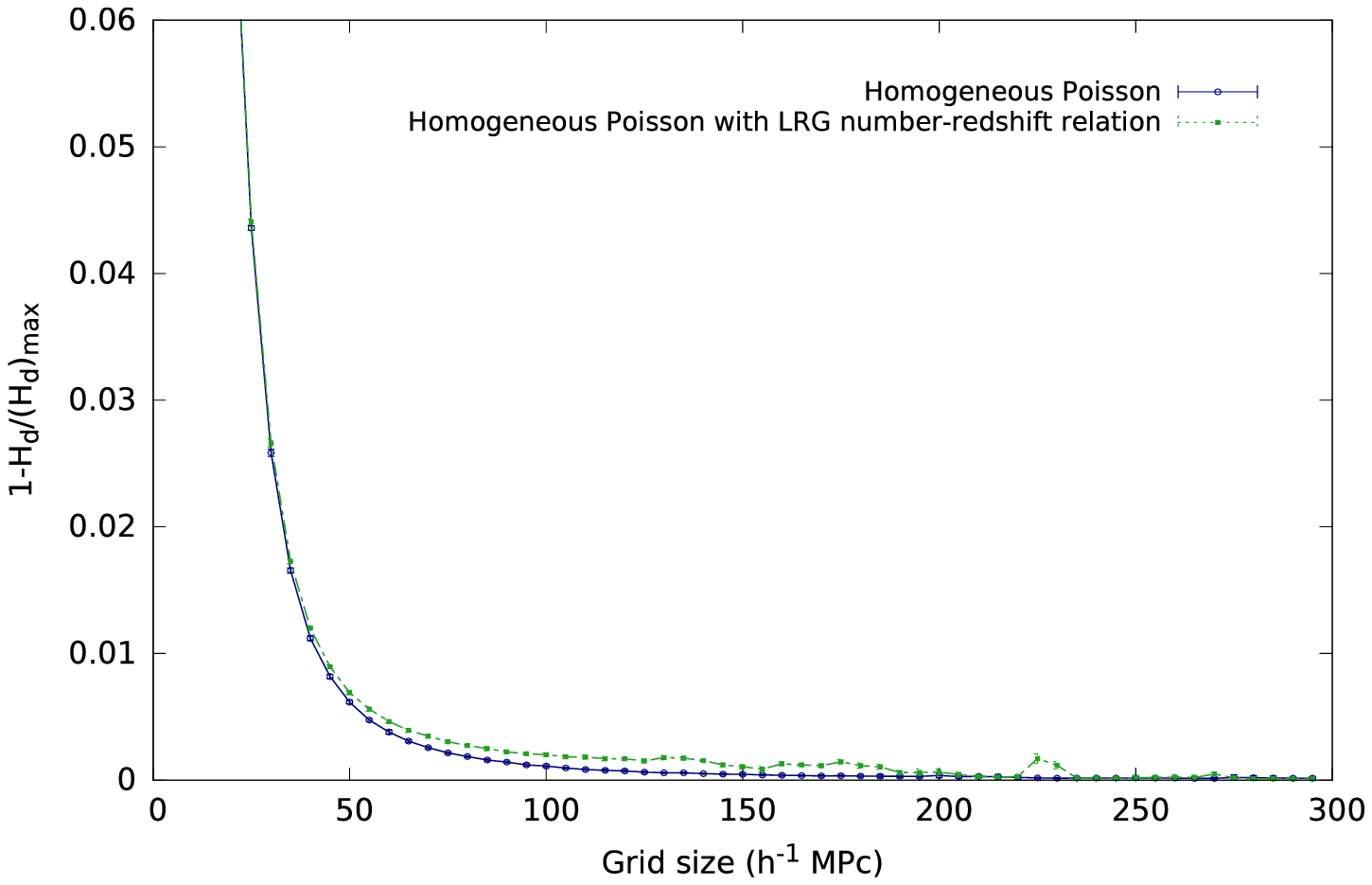}}}%
\resizebox{9cm}{!}{\rotatebox{0}{\includegraphics{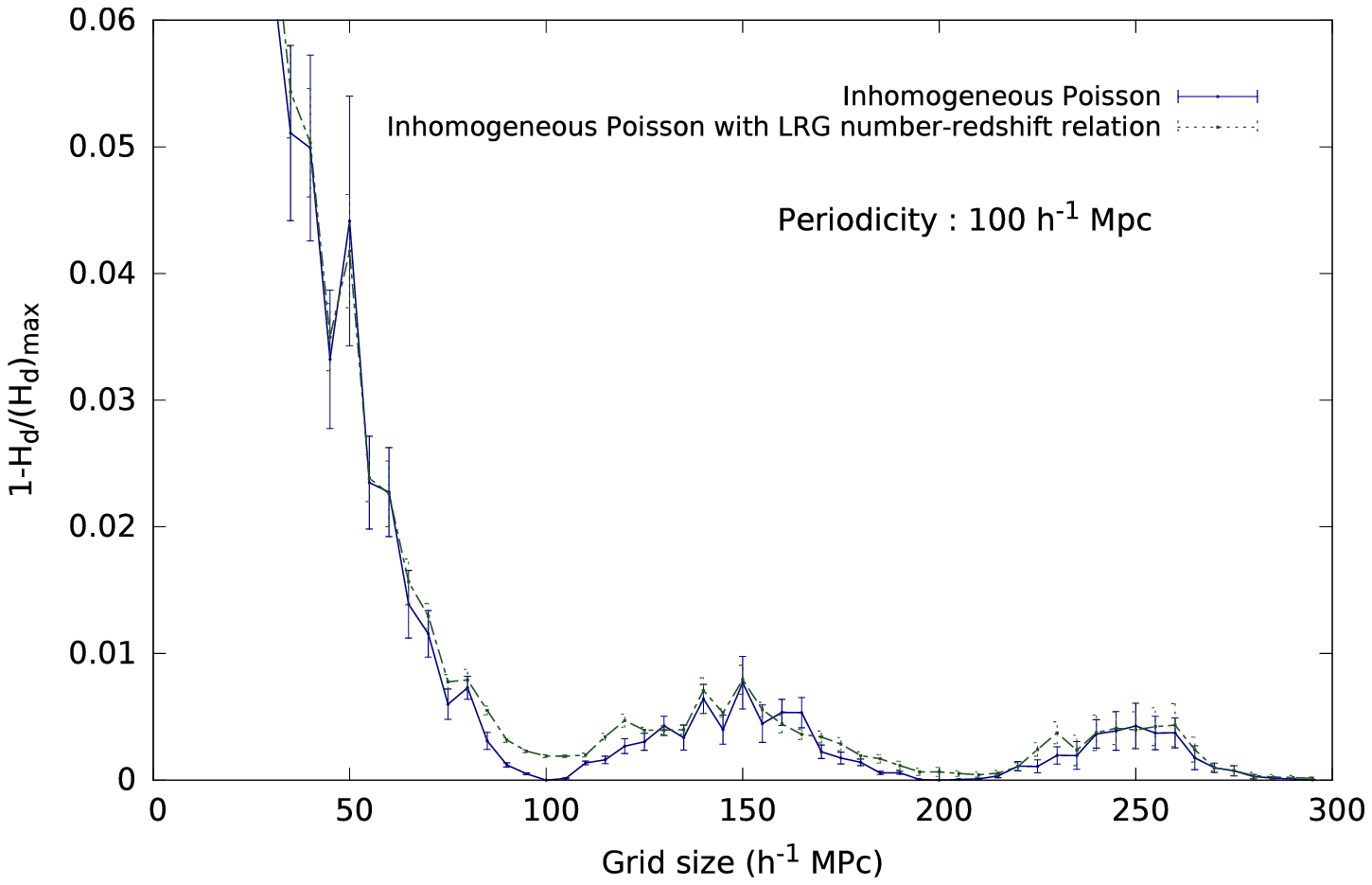}}}\\
 \caption{ The left panel shows $1-\frac{H_{d}}{(H_{d})_{max}}$ as
   function of length scales in the mock LRG samples drawn from
   homogeneous Poisson distributions by enforcing the same comoving
   number density in them as the actual LRG sample. The results for
   mock samples without such restrictions are also shown together. The
   right panels shows the same but for anisotropic and inhomogeneous
   Poisson distributions which are periodic on scales of $100 \,
   h^{-1} \, {\rm Mpc}$. The error-bars shown here in all the panels
   are the $1-\sigma$ variations from the $10$ Monte Carlo
   realizations used in each case.}
  \label{fig:testfig}
\end{figure*}

\section {DATA}
\subsection{LRG DATA}

The Sloan Digital Sky Survey (SDSS) \citep{york} is a wide-field
imaging and spectroscopic survey of the sky using a dedicated 2.5 m
telescope \citep{gunn} with $3^{\circ}$ field of view at Apache Point
Observatory in southern New Mexico. The SDSS has imaged the sky in 5
passbands, $u,\; g,\; r,\; i$ and $z$ covering $10^4$ square degrees
and has so far made 12 data releases to the community. The SDSS galaxy
sample can be roughly divided into two (i) the MAIN galaxy sample and
(ii) the Luminous Red Galaxy (LRG) sample. For our work, we have used
the spectroscopic sample of LRGs derived from the Data Release 7 of
SDSS-II \citep{abaz}.

SDSS targets those galaxies for spectroscopy which have $r$ magnitude
brighter than 17.77 ($r < 17.77$). To select LRGs, additional galaxies
are targeted using color-magnitude cuts in $g,\; r$ and $i$ which
extends the magnitude limit for LRGs to $r < 19.5$. The prominent
feature for an early-type galaxy is the $4000 \, A^{\circ}$ break in
its SED. For $z \lesssim 0.4$ this feature lies in the $g$ passband
while it shifts to the $r$ band for higher redshifts. Hence selection
of LRGs involves different selection criteria below and above $z
\lesssim 0.4$, the details of which are described in \citet{eisen}.

The criteria for lower redshifts are collectively called Cut-I
(eq. (4-9) in \citet{eisen}) while those for the higher redshifts are
called Cut-II (eq.  (9-13) in \citet{eisen}), with Cut-I accounting
for $\approx 80\%$ of the targeted LRGs. A galaxy that passes either
of these cuts is flagged for spectroscopy by the SDSS pipeline as
TARGET\_GALAXY\_RED while a galaxy that passes only Cut II is flagged
as TARGET\_GALAXY\_RED\_II. Hence, while selecting the LRGs we require
that the TARGET\_GALAXY\_RED and TARGET\_GALAXY\_RED\_II flags both be
set.

We obtain the $g$-band absolute magnitude from the $r$ band apparent
magnitude accounting for the k-correction and passive evolution. We
use the prescription for $K+e$ correction from the Table 1 in
\citet{eisen} (non-star forming model). The cuts defined above are
designed to produce an approximately volume limited sample of LRGs
upto $z \approx 0.4$. The comoving number density of LRGs falls
sharply beyond $z \approx 0.4$. It is obviously necessary to include
LRGs that come from the MAIN sample for $z \lesssim 0.3$. But
\citet{eisen} issue a strong advisory against selecting LRGs with $z
\lesssim 0.15$ for a volume limited sample as the luminosity threshold
is not preserved for low redshifts. We have restricted our LRG sample
to the redshift range $0.16 < z < 0.38$ and $g$-band absolute
magnitude range $-23<M_g<-21$. We identify a region $-50<\lambda<50$
and $-33.5<\eta<29.5$ where $\lambda$ and $\eta$ are the survey
coordinates. Combining these cuts provide us our LRG sample which
radially extends from $462 \, h^{-1} {\rm Mpc}$ to $1037 \, h^{-1}
{\rm Mpc}$ and consists of $50513$ luminous red galaxies. We have used
the same data set in an earlier analysis of filamentarity of LRGs in
\citet{pandey11}. The comoving number density of LRGs as a function of
the radial distance $r$ in this sample is shown in Figure 1 of
\citet{pandey11}.

\subsection{MONTE CARLO SIMULATIONS}

 We would like to test if our method can quantify the inhomogeneity
 and characterize the periodicity in a given spatial distribution.  We
 generate a set of Monte Carlo realizations for an anisotropic
 distribution with periodicity in their spatial distribution on
 certain length scale. We decide here to generate the Monte Carlo
 realizations for a simple radial density distributions $\rho(r) = K
 \, \lambda(r)$ where $\lambda(r) = \frac{1}{r^{2}}$. Here $K$ is a
 normalization constant. These distributions are radially
 inhomogeneous Poisson distributions which are anisotropic at all
 points other than the centre.

Enforcing the desired number of particles $N$ within radius $R$ one
can turn the radial density function into a probability function within
$r=0$ to $r=R$ which is normalized to one when integrated over that
interval. So the probability of finding a particle at a given radius
$r$ is $P(r)=\frac{ r^2 \lambda(r) } {\int_{0}^{R} r^2 \lambda(r) \,
  dr}$ which is proportional to the density at that radius implying
more particles in high density regions.

We generate the Monte Carlo realizations of this distribution using a
Monte Carlo dartboard technique. The value of the function $r^2
\lambda(r)$ in $P(r)$ is same and constant everywhere. We label the
maximum value of $P(r)$ as $P_{max}$ which is same everywhere in this
case. We randomly choose a radius $r$ in the range $0 \le r \le R$ and
a probability value is randomly chosen in the range $0 \le P(x) \le
P_{max}$. The actual probability of finding a particle at the selected
radius is then calculated using expression for $P(r)$ and compared to
the randomly selected probability value. If the random probability is
less than the calculated value, the radius is accepted and assigned
isotropically selected angular co-ordinates $\theta$ within $0^{\circ}
\leq \theta \leq 90^{\circ}$ and $\phi$ within $0^{\circ} \leq \phi
\leq 90^{\circ}$, otherwise the radius is discarded. In this way,
radii at which particle is more likely to be found will be selected
more often because the random probability will be more frequently less
than the calculated actual probability. The above $\theta$ and $\phi$
ranges are chosen so as to generate the distribution inside a cubic
region. We choose $R=\sqrt{3} L \, h^{-1} \, {\rm Mpc}$ where $L$ is
the length of the cube on each side. The value of $N$ is chosen
suitably for different $L$ so as to maintain a density which is higher
than the mean density of LRGs. We identify all the points whose
co-ordinates $(x,y,z)$ lie within $0 \leq x \leq L \, h^{-1} \, {\rm
  Mpc}$, $0 \leq y \leq L \, h^{-1} \, {\rm Mpc}$ and $0 \leq z \leq L
\, h^{-1} \, {\rm Mpc}$. This gives us a cube of side $ L \, h^{-1} \,
   {\rm Mpc}$ containing a certain number of points. We replicate this
   cube along x,y and z directions as many times as required so as to
   enable us to extract a region from this periodic distribution which
   has identical geometry as the LRG distribution. We then randomly
   select as many points from the extracted region as there are LRGs
   in our sample.

We choose three different values of $L=50 \, h^{-1} \, {\rm Mpc}$, $L=
100 \, h^{-1} \, {\rm Mpc}$ and $L=150 \, h^{-1} \, {\rm Mpc}$ and generate
$10$ Monte Carlo realizations of the periodic distributions for each
value of $L$. We analyze each set of Monte Carlo realizations
separately using the method described in section 2.

A radially inhomogeneous distribution with no preferred scale must
appear inhomogeneous on all scales and we want to make sure that the
method proposed here efficiently capture the inhomogeneous character
of such distribution on all scales. We generate a radially
inhomogeneous Poisson distribution in a large cubic box where density
falls of as $\frac{1}{r^2}$ from one of the corners of the box. The
total number of points inside the box is decided so as to maintain a
mean density comparable to the mean density of the LRG
distribution. $10$ such boxes are generated using the Monte Carlo
technique discussed above and analyzed separately using the method
presented in section 2.

\section{RESULTS AND CONCLUSIONS}

We show the inhomogeneity $1-\frac{H_{d}}{(H_{d})_{max}}$ as a
function of grid sizes or length scales for the radially inhomogeneous
Poisson distributions in \autoref{fig:inhfig}. The
\autoref{fig:inhfig} shows that the distribution is inhomogeneous on
small scales which is partly due to its inherent inhomogeneous
character and partly due to the Poisson noise. The observed
inhomogeneity decreases with increasing grid sizes on small scales but
suddenly starts increasing at $\sim 50 \, h^{-1} \, {\rm Mpc}$ and keep
on increasing thereafter. The contribution of Poisson noise to the
measured inhomogeneity decreases with increasing grid sizes due to the
increase in the number counts inside them and becomes increasingly
irrelevant on larger scales. The degree of inhomogeneity on larger
scales is primarily decided by the nature of the intrinsic
inhomogeneity built in the distribution. We see that for the grid
sizes above $50 \, h^{-1} \, {\rm Mpc}$ the distribution appears to be
increasingly inhomogeneous. This indicates that there is no scale of
transition to homogeneity in the distribution and the distribution can
be regarded as intrinsically inhomogeneous. It may be noted here that
Figure 2 of \citet{pandey} show that the measurement with overlapping
spheres leads to suppression in inhomogeneity on large scales which
may hide the real inhomogeneous nature of a distribution.  The method
which employs independent voxels is far more reliable in a sense that
there are no artificial reduction in inhomogeneity due to confinement
and overlapping biases. It is this advantage which allow us to capture
the truly inhomogeneous nature of a distribution as shown in
\autoref{fig:inhfig}. The method presented here is certainly
advantageous in this sense and may help us to reliably measure the
inhomogeneity present in a distribution on different scales.

In different panels of \autoref{fig:pinhfig} we show how the
inhomogeneities vary with length scales for inhomogeneous and
anisotropic distributions which have a periodicity in their spatial
distributions (\autoref{fig:dist}). The variations of inhomogeneity in
a homogeneous Poisson distribution is also shown together in each
panel for comparison. In the top left panel we show the inhomogeneity
$1-\frac{H_{d}}{(H_{d})_{max}}$ as a function of grid size for the
distributions which are periodic on scales of $50 \, h^{-1} \, {\rm
  Mpc}$. We see that the periodic distributions and homogeneous
Poisson distributions both show inhomogeneity on small scales which
decrease with increasing length scales. But they vary differently as
the degree of inhomogeneity on small scales would depend on the
Poisson noise as well as any intrinsic inhomogeneity built in the
character of the distributions. It is interesting to note that there
is a sharp decrease in the inhomogeneity in the periodic distributions
at $50 \, h^{-1} \, {\rm Mpc}$. We get
$1-\frac{H_{d}}{(H_{d})_{max}}=0$ when the grid size is $50 \, h^{-1}
\, {\rm Mpc}$ which also is the periodicity of the distributions. The
inhomogeneity reappears again as the grid size exceeds $50 \, h^{-1} \,
{\rm Mpc}$ and decreases once again and disappears exactly at $100 \,
h^{-1} \, {\rm Mpc}$ and so on. This can be easily explained from the
fact that these periodic distributions are highly inhomogeneous below
the scale of their periodicity and become homogeneous at the scale at
which they are repeated. Once the grid size surpass this scale the
distribution would again appear inhomogeneous by partial inclusion of
the repeated patterns by the voxels. This behaviour of
$1-\frac{H_{d}}{(H_{d})_{max}}$ continues with increasing grid sizes
and we get successive minima at repeated intervals of $50 \, h^{-1} \,
{\rm Mpc}$ showing precise homogeneity on those length scales. The
maximum amplitude of the inhomogeneity decreases in successive
oscillations due to the decrease in the Poisson noise with increasing
grid sizes. The error-bars shown here are the $1-\sigma$ variations
from the $10$ Monte Carlo realizations used in each case. The
extremely small size of the errorbars on the measured inhomogeneity at
each node clearly indicates the presence of a periodicity in the
distributions. This characteristic behaviour of inhomogeneity in
periodic distributions is markedly different from the smooth decrease
of inhomogeneity with increasing length scales in the homogeneous
Poisson distributions.  We do not find this periodic variation in the
homogeneous Poisson distributions as there are no characteristic
scales associated with them. The variations of
$1-\frac{H_{d}}{(H_{d})_{max}}$ as a function of grid size for
periodic distributions with periodicity of $100 \, h^{-1} \, {\rm
  Mpc}$ and $150 \, h^{-1} \, {\rm Mpc}$ are shown in the top right
and bottom panels respectively. We see similar periodic variations in
the measured inhomogeneity at repeated intervals of $100 \, h^{-1} \,
{\rm Mpc}$ and $150 \, h^{-1} \, {\rm Mpc}$ in the respective panels
which clearly demonstrate the presence of periodicity in the
distributions on those scales. This shows that our method can identify
the presence of any regularity in a spatial distribution and quantify
the scale over which the patterns are repeated.

We show the inhomogeneity $1-\frac{H_{d}}{(H_{d})_{max}}$ in the LRG
distribution and a homogeneous Poisson distribution as a function of
length scales in the top left panel of \autoref{fig:lrghfig} when the
data is analyzed with independent voxels. We find that both
distributions show inhomogeneity on small scales but the degree of
inhomogeneity observed in the LRG distribution is higher than that of
a homogeneous Poisson distribution at each length scales. This
indicates that the LRGs are not distributed randomly and the amount of
information encoded in the LRG distribution is higher as compared to a
homogeneous Poisson distribution at each scale. The degree of
inhomogeneity in both the LRG and Poisson distribution decrease with
increasing length scales but the inhomogeneity in the Poisson
distribution decays much faster as compared to the LRG
distribution. This can be explained from the fact that the information
in the LRG distribution is sourced by the gravitational clustering and
Poisson noise whereas for the Poisson distribution the contribution
solely comes from the shot noise which becomes increasingly less
important on larger length scales. In the top right panel of
\autoref{fig:lrghfig} we show the inhomogeneity
$1-\frac{H_{r}}{(H_{r})_{max}}$ in the LRG distribution as a function
of length scales when the analysis is carried out with overlapping
spheres. We find that the distribution appears to be homogeneous on
small scales. The degree of inhomogeneity suddenly rises at a certain
length scale after which it gradually decreases with increasing length
scales as seen earlier in the top left panel of
\autoref{fig:lrghfig}. This characteristic behaviour giving rise to a
bump in the inhomogeneity at a given scale arise due to the near
uniformity in the distribution on length scale below its mean inter
particle separation. It may be noted that the peak of the bump is
located at $\sim 23 \, h^{-1} \, {\rm Mpc}$ which is the mean
intergalactic separation in our LRG sample. A Poisson distribution is
expected to become homogeneous on a scale of the average size of voids
in the distribution \citep{sdm}. This is closely related to the mean
inter-particle separation and one would expect a Poisson distribution
to become homogeneous above this length scale. But we find a large
degree of inhomogeneity in the LRG distribution on these length
scales. Further we notice that the degree of inhomogeneity in the top
right panel is smaller than that in the top left panel at each length
scales. These differences are caused by the suppression of
inhomogeneities due to confinement and overlapping biases
\citep{pandey,kraljic} which result from the finite volume and
progressive overlaps between the measuring spheres with increasing
radii. In the bottom left panel of \autoref{fig:lrghfig} we show the
ratio of inhomogeneities measured at different length scales using
independent voxels and overlapping spheres. We find that the
inhomogeneities are suppressed by a factor of $\sim 5$ upto $170 \,
h^{-1} \, {\rm Mpc}$ and by a factor of $10$ on even larger scales in
the overlapping measurements. The ratio shows a sudden drop at $\sim
200 \, h^{-1} \, {\rm Mpc}$ due to the unavailability of sufficient
independent voxels beyond this length scale. The results clearly
demonstrate the suppression of inhomogeneities on larger length scales
when the measuring spheres are allowed to overlap. All the methods for
testing homogeneity which are based on the number counts $n(<r)$ in
spheres of radius $r$ are equally affected by these biases.  Therefore
these biases should be properly taken into account in their
applications.

In the left panel of \autoref{fig:lrgmrfig} we show how the number of
available voxels vary with length scales when independent voxels are
used for the analysis. Clearly the number of voxels decreases with
increasing grid sizes due to the finite volume of the survey
region. We see that the number of independent voxels drops below $10$
around $\sim 200 \, h^{-1} \, {\rm Mpc}$ and we decide not to consider
the scales beyond this in our analysis. The right panel of
\autoref{fig:lrgmrfig} show the number of available centres as a
function of the radius of the spheres when overlapping spheres are
used for the analysis. Clearly in this method one can go upto a larger
length scales albeit by compromising with larger degree of overlaps
between the measuring spheres resulting into a greater degree of
suppression in inhomogeneity. We see that one can extend this analysis
upto $\sim 270 \, h^{-1} \, {\rm Mpc}$ requiring at least $\sim 100$
overlapping spheres. It may be noted in the top right panel of
\autoref{fig:lrghfig} that $1-\frac{H_{r}}{(H_{r})_{max}}$ decreases
to $10^{-3}$ at $150 \, h^{-1} \, {\rm Mpc}$ and show very little
variation thereafter despite the increasing degree of overlaps between
the measuring spheres. The value of $1-\frac{H_{d}}{(H_{d})_{max}}$ in
the top left panel of \autoref{fig:lrghfig} also exhibit a quite
similar behaviour where its value decreases to $5 \times 10^{-3}$ at
$150 \, h^{-1} \, {\rm Mpc}$ and show a flattening which abruptly
drops to $\approx 0$ at $\sim 200 h^{-1} \, {\rm Mpc}$. We do not
attach any statistical significance to this sudden drop due to the non
availability of sufficient number of voxels on that scale. The degree
of inhomogeneity at the plateau is somewhat different in the two
methods. This difference could be attributed to the fact that
inhomogeneities are suppressed by a factor of $\sim 5$ at $150 \,
h^{-1} \, {\rm Mpc}$ when overlapping spheres are used
(\autoref{fig:lrghfig}). We note that $\sim 100$ independent voxels
and $>10^{4}$ centers of overlapping spheres are available at $150 \,
h^{-1} \, {\rm Mpc}$ which trace a very small degree of inhomogeneity
present on that scale. The minuscule degree of inhomogeneity measured
with a large number of voxels or spheres at the plateau indicates that
the LRG distribution is nearly homogeneous on these length scales. The
LRG distribution is known to have a linear bias of $\sim 2$ on large
scales \citep{martin, sawang} and may have a larger variations in
$1-\frac{H_{d}}{(H_{d})_{max}}$ as compared to the homogeneous Poisson
distributions. Currently we do not have any estimates for the
variations in $1-\frac{H_{d}}{(H_{d})_{max}}$ as we have only one
distribution for the LRGs. Keeping this in mind we conclude that the
LRG distribution becomes homogeneous on scales of $150 \, h^{-1} \,
{\rm Mpc}$ and beyond. Our results are consistent with the earlier
findings that the distribution of galaxies in the SDSS Main sample
becomes homogeneous on and above a scale of $140 \, h^{-1} \, {\rm
  Mpc}$ \citep{pandey15}. In the bottom right panel of
\autoref{fig:lrghfig} we show the Kullback-Leibler(KL) divergence or
the information divergence for the LRG distribution with respect to a
homogeneous Poisson distribution on different length scales when the
measurements are done with independent voxels. The definition and the
significance of this measure is discussed in section 2 and this can be
regarded as an alternative measure of inhomogeneity. It is interesting
to note that the KL divergence also decreases with increasing grid
sizes which is quite similar to the decrease in
$1-\frac{H_{d}}{(H_{d})_{max}}$ and $1-\frac{H_{r}}{(H_{r})_{max}}$
with increasing length scales and a similar flattening is observed at
$\sim 150 \, h^{-1} \, {\rm Mpc}$.

An important caveat to the present analyses arises due the fact that
the comoving number density of LRGs varies with redshift and the LRG
sample is considered to be only quasi volume limited
\citep{eisen,zehavi,kazin}. These variations could result from both
the intrinsic fluctuations in the galaxy density field and the
observational selection effects \citep{labini}. We test how the
presence of such fluctuations in number-redshift relation affect the
scale of homogeneity inferred with our method. To test this we
construct $10$ mock LRG samples each from homogeneous and
inhomogeneous Poisson distributions and randomly exclude points from
them so as to finally have the same comoving number density $n(z)$ in
them as the actual LRG sample. The results for these analyses are
shown in \autoref{fig:testfig}. In two panels of \autoref{fig:testfig}
we see a small increase in the degree of inhomogeneity when an
LRG-like number-redshift relation is enforced in the homogeneous and
inhomogeneous Poisson distributions. But we note that the variations
of inhomogeneity with length scales in both cases still reflect the
characteristics of the underlying distributions and the respective
scales of transition to homogeneity in these distributions are not
altered by the presence of such a number-redshift relation. However it
may be worth mentioning that this test and its outcome rely on a
smooth variation in the radial density due to either selection effects
or intrinsic fluctuations in the galaxy density field and it is in
general difficult to test for any other sources of such fluctuations
in the number-redshift relation of LRGs.

We do not see any periodic variation in the measured inhomogeneity
$1-\frac{H_{d}}{(H_{d})_{max}}$ of the LRG distribution
(\autoref{fig:lrghfig}) in the range of length scales explored. This
is in sharp contrast with the variations observed in
\autoref{fig:pinhfig} for periodic distributions. This indicates that
there are no underlying regularities in the distribution of the
luminous red galaxies at least within the length scales probed in the
present work. One can extend this analysis to the LRG sample from the
Baryon Oscillation Spectroscopic Survey of SDSS-III \citep{dawson} and
future LRG sample such as from the eBOSS of SDSS-IV \citep{prakash} to
search for any periodic variation in inhomogeneity on even larger
scales.

\section{ACKNOWLEDGEMENT}
The authors would like to thank an anonymous reviewer for useful
comments and suggestions. The authors thank Prajwal Raj Kafle for his
suggestion for using KL divergence as an alternative measure of
inhomogeneity. B.P. would like to acknowledge IUCAA, Pune for
providing support through the Associateship Programme. B.P. would also
like to acknowledge CTS, IIT Kharagpur for the use of its facilities
for the present work.

Funding for the creation and distribution of the SDSS Archive has been 
provided by the Alfred P. Sloan Foundation, the Participating 
Institutions, the National Aeronautics and Space Administration, the 
National Science Foundation, the U.S. Department of Energy, the Japanese 
Monbukagakusho, and the Max Planck Society. The SDSS Web site is 
http://www.sdss.org/.

    The SDSS is managed by the Astrophysical Research Consortium (ARC) for 
the Participating Institutions. The Participating Institutions are The 
University of Chicago, Fermilab, the Institute for Advanced Study, the 
Japan Participation Group, The Johns Hopkins University, the Korean 
Scientist Group, Los Alamos National Laboratory, the Max-Planck-Institute 
for Astronomy (MPIA), the Max-Planck-Institute for Astrophysics (MPA), New 
Mexico State University, University of Pittsburgh, Princeton University, 
the United States Naval Observatory, and the University of Washington.

\bsp	
\label{lastpage}
\end{document}